\documentclass[12pt]{article}
\usepackage{amssymb,graphicx}

\newcommand{\Tr}{\mathop{\mathrm{Tr}}\nolimits}
\renewcommand{\slash}[1]{#1\llap{/}}
\newcommand{\sslash}[1]{#1\llap{$\scriptstyle/$}}
\newcommand{\LMS}{\Lambda_{\overline{\mathrm{MS}}}}
\newcommand{\ep}{\varepsilon}
\newcommand{\tfrac}[2]{{\textstyle\frac{#1}{#2}}}

\begin{document}
{\Large\textbf{Recent progress on HQET lagrangian}}\\[2mm]
\textit{A.~G.~Grozin}\\[2mm]
{\small Budker Institute of Nuclear Physics, Novosibirsk 630090, Russia}\\[2mm]
{\small HQET lagrangian up to $1/m^3$ terms is discussed.
Consequences of reparameterization invariance are considered.
Results for the chromomagnetic interaction coefficient
at two loops, and in all orders in the large--$\beta_1$ approximation,
are presented.}

\section{HQET lagrangian}

QCD problems with a single heavy quark staying approximately at rest
can be conveniently treated in the heavy quark effective field theory
(HQET) (see~\cite{Neubert} for review and references).
We shift the energy zero level: $E=m+\omega$,
and consider the region where residual energies $\omega$ and momenta $\vec{p}$
are not large: $\omega\sim|\vec{p}|\sim\Lambda\ll m$.
The effective field theory is constructed to reproduce
QCD on--shell scattering amplitudes expanded to some order $(\Lambda/m)^n$.
This is achieved by writing down the most general effective Lagrangian
consistent with the required symmetries,
and tuning the coefficients to reproduce QCD on-shell amplitudes.
Terms with $D_0 Q$ can be eliminated by field redefinitions.

The most general lagrangian up to $1/m^3$ is~\cite{EH1}--\cite{Manohar}
\begin{eqnarray}
&&\hspace{-6mm}L = Q^+ i D_0 Q
\nonumber\\
&&\hspace{-6mm} + \frac{C_k}{2m} Q^+ \vec{D}^2 Q
+ \frac{C_m}{2m} Q^+ \vec{B}\cdot\vec{\sigma} Q
+ \frac{i C_s}{8m^2} Q^+ (\vec{D}\times\vec{E}-\vec{E}\times\vec{D})
\cdot\vec{\sigma} Q
+ \frac{C_d}{8m^2} Q^+ [\vec{D}\cdot\vec{E}] Q
\nonumber\\
&&\hspace{-6mm} + \frac{C_{k2}}{8m^3} Q^+ \vec{D}^4 Q
+ \frac{C_{w1}}{8m^3} Q^+ \{\vec{D}^2,\vec{B}\cdot\vec{\sigma}\} Q
- \frac{C_{w2}}{4m^3} Q^+ D^i \vec{B}\cdot\vec{\sigma} D^i Q
\label{l0}\\
&&\hspace{-1mm} + \frac{C_{p'p}}{8m^3} Q^+ (\vec{D} \vec{B}\cdot\vec{D}
  +\vec{D}\cdot\vec{B} \vec{D}) \cdot\vec{\sigma} Q
+ \frac{i C_M}{8m^3} Q^+ (\vec{D}\cdot[\vec{D}\times\vec{B}]
  + [\vec{D}\times\vec{B}]\cdot\vec{D}) Q
\nonumber\\
&&\hspace{-1mm} + \frac{C_{a1}}{8m^3} Q^+ (\vec{B}^2-\vec{E}^2) Q
- \frac{C_{a2}}{16m^3} Q^+ \vec{E}^2 Q
+ \frac{C_{a3}}{8m^3} Q^+ \Tr(\vec{B}^2-\vec{E}^2) Q
- \frac{C_{a4}}{16m^3} Q^+ \Tr\vec{E}^2 Q
\nonumber\\
&&\hspace{-1mm} + \frac{i C_{b1}}{8m^3} Q^+ (\vec{B}\times\vec{B}
  -\vec{E}\times\vec{E}) \cdot\vec{\sigma} Q
- \frac{i C_{b2}}{8m^3} Q^+ (\vec{E}\times\vec{E}) \cdot\vec{\sigma} Q
+ \cdots
\nonumber
\end{eqnarray}
where $Q$ is 2--component heavy--quark field.
Here heavy--light contact interactions are omitted,
as well as operators involving only light fields.

HQET can be rewritten in relativistic notations.
Momenta of all states are decomposed as $p=mv+k$
where residual momenta $k\sim\Lambda$.
The heavy--quark field is now Dirac spinor obeying $\slash{v}Q_v=Q_v$.
The lagrangian is
\begin{eqnarray}
&&\hspace{-6mm}
L_v = \overline{Q}_v i v\cdot D Q_v
- \frac{C_k}{2m} \overline{Q}_v D_\bot^2 Q_v
- \frac{C_m}{4m} \overline{Q}_v G_{\mu\nu}\sigma^{\mu\nu} Q_v
\label{l1}\\
&&\hspace{-6mm}
+ \frac{i C_s}{8m^2}
\overline{Q}_v \{D_\bot^\mu,G^{\lambda\nu}\}v_\lambda \sigma_{\mu\nu} Q_v
- \frac{C_d}{8m^2} \overline{Q}_v v^\mu [D_\bot^\nu G_{\mu\nu}] Q_v
+ \cdots
\nonumber
\end{eqnarray}
where $D_\bot=D-v(vD)$.
The velocity $v$ may be changed by an amount $\delta v\lesssim\Lambda/m$
without spoiling the applicability of HQET and changing its predictions.
This reparameterization invariance relates coefficients of varying degrees
in $1/m$~\cite{LM}--\cite{Lee3}.

At the tree level, there are easier ways to find the coefficients $C_i$
than QCD/HQET matching: Foldy--Wouthuysen transformation~\cite{KT,BKP},
or using equations of motion~\cite{Lee}
(or integrating out lower components~\cite{MRR,Lee2})
followed by a field redefinition.
The result is
\begin{eqnarray}
&&C_k=C_m=C_d=C_s=C_{k2}=C_{w1}=C_{a1}=C_{b1}=1\,,
\label{tree}\\
&&C_{w2}=C_{p'p}=C_M=C_{a2}=C_{a3}=C_{a4}=C_{b2}=0\,.
\nonumber
\end{eqnarray}
However, these algebraic methods don't generalize to higher loops.

At $1/m$ level, the kinetic coefficient $C_k=1$
due to the reparameterization invariance~\cite{LM}.
One--loop matching for the chromomagnetic coefficient $C_m$
was done in~\cite{EH2};
two--loop anomalous dimension of the chromomagnetic operator in HQET
was obtained in~\cite{ABN,CG}, and two--loop matching was done in~\cite{CG};
in~\cite{GN}, all orders of perturbation theory for $C_m$ were summed
at large $\beta_1$.

At $1/m^2$ level, the spin--orbit coefficient $C_s=2C_m-1$
due to the reparameterization invariance~\cite{CKO}--\cite{BKPR}.
The Darwin term reduces to a contact interaction.
One--loop matching for the heavy--light contact interactions
was done in~\cite{BKPR}.
The one--loop anomalous dimension matrix of dimension 6 terms
in the HQET lagrangian was obtained in~\cite{BKP}, \cite{BO}--\cite{BM}.

At $1/m^3$ level, one--loop matching was done in~\cite{Manohar}
for the terms involving the heavy--quark fields twice and the gluon field once.
The one--loop renormalization of dimension 7 terms
in the HQET lagrangian was recently considered~\cite{Balzereit2}.

\section{Matching quark--quark vertex}

Renormalized QCD on--shell quark--quark proper vertex
\begin{equation}
-\overline{u}(\slash{p}-m)u
\label{QCD2}
\end{equation}
gets no correction in the on--shell renormalization scheme.
QCD spinors are related to HQET spinors
by the Foldy--Wouthuysen transformation
\begin{equation}
u=\left(1+\frac{\slash{k}}{2m}+\frac{k^2}{4m^2}+\cdots\right)u_v\,,\quad
\slash{v}u_v=u_v\,.
\label{FW}
\end{equation}
Expressing QCD proper vertex via HQET spinors, we obtain
\begin{equation}
\overline{u}_v \frac{\vec{k}^2}{2m} u_v + \cdots
\label{QCD2h}
\end{equation}

Let's denote the sum of bare 1--particle--irreducible self--energy diagrams
of the heavy quark in HQET at $1/m^0$
as $-i\frac{1+\sslash{v}}{2}\Sigma(\omega)$, $\omega=kv$.
At the $1/m$ level, self--energy diagrams
with a single chromomagnetic vertex vanish.
Let the sum of bare diagrams with a single kinetic vertex be
$-i\frac{C_k}{2m}\frac{1+\sslash{v}}{2}\Sigma_k(\omega,k_\bot^2)$.
Consider variation of $\Sigma$ at $v\to v+\delta v$
for an infinitesimal $\delta v$ ($v\,\delta v=0$).
All factors $\frac{1+\sslash{v}}{2}$ can be combined into a single one,
and the variation $\delta\slash{v}$ in it provides the variation
of the $\gamma$--matrix structure in front of $\Sigma$.
There are two sources of the variation of $\Sigma$.
Terms from the expansion of denominators of the propagators
produce insertions $ik\delta v$.
Terms from the vertices produce $igt^a\delta v^\mu$.
Now consider variation of $\Sigma_k$ at $k_\bot\to k_\bot+\delta k_\bot$
for an infinitesimal $\delta k_\bot$.
Quark--quark kinetic vertices produce $i\frac{C_k}{m}k\delta k_\bot$;
quark--quark--gluon kinetic vertices
produce $i\frac{C_k}{m}gt^a\delta k_\bot^\mu$;
two--gluon vertices produce nothing.
Therefore,
\begin{equation}
\frac{\partial\Sigma_k}{\partial k_\bot^\mu} =
2 \frac{\partial\Sigma}{\partial v^\mu}\,.
\label{Ward1}
\end{equation}
This is the Ward identity of the reparameterization invariance
first derived in~\cite{Balzereit}.
Taking into account
$\frac{\partial\Sigma_k}{\partial k_\bot^\mu}=
2\frac{\partial\Sigma_k}{\partial k_\bot^2}k_\bot^\mu$
and
$\frac{\partial\Sigma}{\partial v^\mu}=
\frac{d\Sigma}{d\omega}k_\bot^\mu$,
we obtain
\begin{equation}
\frac{\partial\Sigma_k}{\partial k_\bot^2} =
\frac{d\Sigma}{d\omega}\,.
\label{Ward2}
\end{equation}
The right--hand side does not depend on $k_\bot^2$, and hence
\begin{equation}
\Sigma_k(\omega,k_\bot^2) = \frac{d\Sigma(\omega)}{d\omega} k_\bot^2
+ \Sigma_{k0}(\omega)\,.
\label{Ward3}
\end{equation}
This result can also be understood in a more direct way.
Only diagrams with a quark--quark kinetic vertex contain $k_\bot^2$;
its coefficient is is $i\frac{C_k}{2m}$.
The sum of diagrams with a unit insertion is $-i\frac{d\Sigma}{d\omega}$.
Note that diagrams with a quark--quark--gluon kinetic vertex vanish
because there is no preferred transverse direction.

\begin{figure}[ht]
\includegraphics[width=\linewidth]{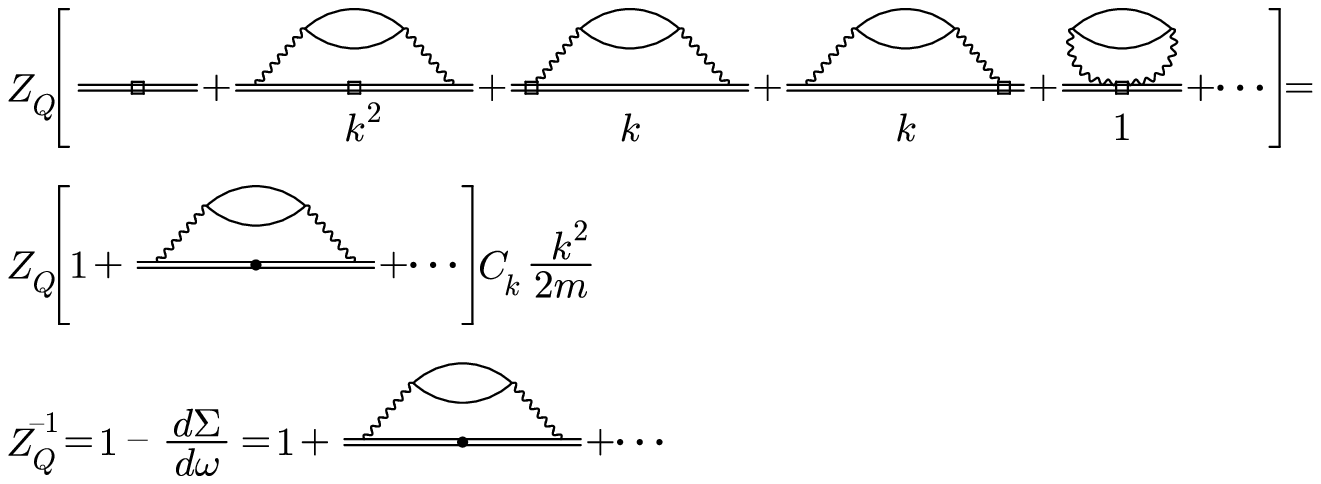}
\caption{HQET quark--quark proper vertex on the mass shell}
\label{Fig:1}
\end{figure}

On the mass shell ($\omega=0$), the renormalized HQET quark--quark
proper vertex is
$\frac{C_k}{2m} Z_Q \overline{u}_v \allowbreak
\bigl[ -k_\bot^2 + \Sigma_k(0,k_\bot^2)
\bigr] u_v = - \frac{C_k}{2m} Z_Q \left[ 1 - \frac{d\Sigma}{d\omega}
\right]_{\omega=0} k_\bot^2 \overline{u}_v u_v$.
On the mass shell, only diagrams with finite--mass particles in loops
contribute (e.g., $c$--quark loops in $b$--quark HQET) (Fig.~\ref{Fig:1}).
Taking into account
$Z_Q^{-1}=1-\left.\frac{d\Sigma}{d\omega}\right|_{\omega=0}$
and comparing with~(\ref{QCD2h}), we finally obtain
\begin{equation}
C_k(\mu)=1\,.
\label{Ck}
\end{equation}
This argument works for an arbitrary $\mu$;
hence, the anomalous dimension of the kinetic--energy operator in HQET
vanishes exactly.
In a similar way, it is not difficult to prove that
\begin{equation}
C_{k2}=1\,.
\label{Ck2}
\end{equation}

\section{Matching quark--quark--gluon vertex}

QCD on--shell proper vertex is characterized by 2 form factors:
\begin{eqnarray}
&&\overline{u}(p') t^a \left( \ep(q^2) \frac{(p+p')^\mu}{2m}
+ \mu(q^2) \frac{[\slash{q},\gamma^\mu]}{4m} \right) u(p)\,,
\label{FF}\\
&&\ep(q^2) = 1 + \ep'\frac{q^2}{m^2} + \cdots, \quad
\mu(q^2) = \mu + \mu'\frac{q^2}{m^2} + \cdots
\nonumber
\end{eqnarray}

\begin{figure}[ht]
\begin{picture}(160,10)
\put(0,0){\includegraphics{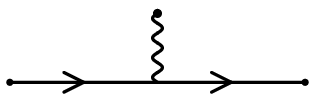}}
\put(8,6){\makebox(0,0){$p$}}
\put(24,6){\makebox(0,0){$p+q$}}
\put(32,2.5){\makebox(0,0)[l]{${}=g\,e^a t^a$}}
\put(56,2.5){\makebox(0,0)[l]{$\Biggl[\Biggr.$}}
\put(57,0){\includegraphics{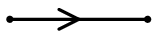}}
\put(65,6){\makebox(0,0){$p+q$}}
\put(75,2.5){\makebox(0,0){$-$}}
\put(77,0){\includegraphics{A1.eps}}
\put(85,6){\makebox(0,0){$p$}}
\put(93,2.5){\makebox(0,0)[l]{$\Biggl.\Biggr]=g\Biggl[\Biggr.$}}
\put(104,0){\includegraphics{A1.eps}}
\put(113,6){\makebox(0,0){$p+q e^a t^a$}}
\put(122,2.5){\makebox(0,0){$-$}}
\put(124,0){\includegraphics{A1.eps}}
\put(132,6){\makebox(0,0){$p$}}
\put(140,2.5){\makebox(0,0)[l]{$\Biggl.\Biggr]$}}
\end{picture}
\begin{picture}(160,12)
\put(0,0){\includegraphics{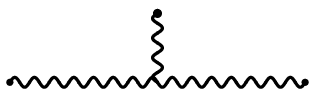}}
\put(8,6){\makebox(0,0){$p$}}
\put(24,6){\makebox(0,0){$p+q$}}
\put(1,1){\makebox(0,0)[t]{$n$}}
\put(31,1){\makebox(0,0)[t]{$m$}}
\put(32,2.5){\makebox(0,0)[l]{${}=g\,e^a (t^a)^{mn}$}}
\put(56,2.5){\makebox(0,0)[l]{$\Biggl[\Biggr.$}}
\put(57,0){\includegraphics{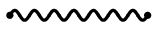}}
\put(65,6){\makebox(0,0){$p+q$}}
\put(75,2.5){\makebox(0,0){$-$}}
\put(77,0){\includegraphics{A3.eps}}
\put(85,6){\makebox(0,0){$p$}}
\put(93,2.5){\makebox(0,0)[l]{$\Biggl.\Biggr]=g\Biggl[\Biggr.$}}
\put(104,0){\includegraphics{A3.eps}}
\put(113,6){\makebox(0,0){$p+q e^a t^a$}}
\put(122,2.5){\makebox(0,0){$-$}}
\put(124,0){\includegraphics{A3.eps}}
\put(132,6){\makebox(0,0){$p$}}
\put(140,2.5){\makebox(0,0)[l]{$\Biggl.\Biggr]$}}
\end{picture}
\begin{picture}(160,18)
\put(0,0){\includegraphics{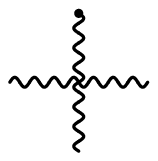}}
\put(1,6){\makebox(0,0){$l$}}
\put(15,6){\makebox(0,0){$n$}}
\put(11,2){\makebox(0,0){$m$}}
\put(16,8.5){\makebox(0,0)[l]{$=g\,e^a\Biggl[\Biggl(\Biggr.\Biggr.$}}
\put(32,0){\includegraphics{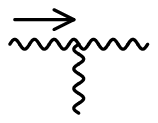}}
\put(36,13){\makebox(0,0)[b]{$+q$}}
\put(33,6){\makebox(0,0){$x$}}
\put(50,8.5){\makebox(0,0){$-$}}
\put(52,0){\includegraphics{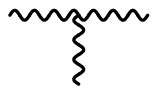}}
\put(53,6){\makebox(0,0){$x$}}
\put(69,8.5){\makebox(0,0)[l]{$\Biggl.\Biggr)(t^a)^{xl}$}}
\end{picture}
\begin{picture}(160,18)
\put(0,8.5){\makebox(0,0)[l]{${}+\Biggl(\Biggr.$}}
\put(7,0){\includegraphics{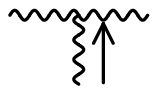}}
\put(18.5,4){\makebox(0,0)[l]{$+q$}}
\put(14,2){\makebox(0,0)[r]{$x$}}
\put(25,8.5){\makebox(0,0){$-$}}
\put(27,0){\includegraphics{A6.eps}}
\put(34,2){\makebox(0,0)[r]{$x$}}
\put(44,8.5){\makebox(0,0)[l]{$\Biggl.\Biggr)(t^a)^{xm} + \Biggl(\Biggr.$}}
\put(63,0){\includegraphics{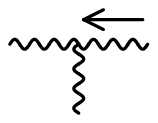}}
\put(75,13){\makebox(0,0)[b]{$+q$}}
\put(78,6){\makebox(0,0){$x$}}
\put(81,8.5){\makebox(0,0){$-$}}
\put(83,0){\includegraphics{A6.eps}}
\put(98,6){\makebox(0,0){$x$}}
\put(100,8.5){\makebox(0,0)[l]{$\Biggl.\Biggr)(t^a)^{xn}\Biggl.\Biggr]$}}
\end{picture}
\begin{picture}(160,18)
\put(0,8.5){\makebox(0,0)[l]{${}=g\Biggl[\Biggr.$}}
\put(10,0){\includegraphics{A5.eps}}
\put(15,13){\makebox(0,0)[b]{$+qet$}}
\put(28,8.5){\makebox(0,0){$-$}}
\put(30,0){\includegraphics{A6.eps}}
\put(47,8.5){\makebox(0,0)[l]{$+$}}
\put(50,0){\includegraphics{A8.eps}}
\put(61.5,4){\makebox(0,0)[l]{$+qet$}}
\put(68,8.5){\makebox(0,0){$-$}}
\put(70,0){\includegraphics{A6.eps}}
\put(87,8.5){\makebox(0,0)[l]{$+$}}
\put(90,0){\includegraphics{A7.eps}}
\put(101,13){\makebox(0,0)[b]{$+qet$}}
\put(108,8.5){\makebox(0,0){$-$}}
\put(110,0){\includegraphics{A6.eps}}
\put(127,8.5){\makebox(0,0)[l]{$\Biggl.\Biggr]$}}
\end{picture}
\caption{Ward identities in the background field formalism}
\label{Fig:2}
\end{figure}

The total colour charge of a quark $\ep(0)=1$ due to the gauge invariance.
Ward identities in the background field formalism~\cite{Abbott}
are shown in Fig.~\ref{Fig:2}, where the large dot means convolution
with the gluon incoming momentum $q$ and colour polarization $e^a$,
the second equalities are valid only for an infinitesimal $q$
(or in the case of an abelian external field),
and $(t^a)^{bc}=if^{acb}$ in the adjoint representation.
Therefore, the QCD proper vertex $\Lambda_\mu^a(p,q)=\Lambda_\mu t^a$
obeys $\Lambda_\mu^a q^\mu e^a=-\Sigma(p+qe^a t^a)+\Sigma(p)$
for infinitesimal $q$, or
$\Lambda_\mu(p,0)=-\frac{\partial\Sigma(p)}{\partial p^\mu}$.
The form factor is projected out by
$\ep(0)=Z_Q\bigl[1+\frac{1}{4}\Tr\Lambda_\mu v^\mu(1+\slash{v})\bigr]$.
On the mass shell,
$\frac{1}{4}\Tr\frac{\partial\Sigma}{\partial p^\mu}=(1-Z_Q^{-1})v_\mu$,
and hence $\ep(0)=1$.

Let's denote the sum of bare vertex diagrams in HQET at $1/m^0$
as $igt^a v^\mu \frac{1+\sslash{v}}{2}[1+\Lambda(\omega,\Delta)]$,
where $\Delta=qv=\omega'-\omega$.
The Ward identity for the static quark propagator is the same
as for the ordinary one (Fig.~\ref{Fig:2}).
Therefore, $\Delta e^a t^a \Lambda(\omega,\Delta)=
-\Sigma(\omega+\Delta e^a t^a)+\Sigma(\omega)$ for infinitesimal $\Delta$, or
\begin{equation}
\Lambda(\omega,0)=-\frac{d\Sigma(\omega)}{d\omega}\,.
\label{Wh}
\end{equation}
It is interesting, that for an abelian external field
$\Lambda(\omega,\Delta)=
-\frac{\Sigma(\omega+\Delta)-\Sigma(\omega)}{\Delta}$
exactly.
The total colour charge of a static quark $Z_Q[1+\Lambda(0,0)]=1$,
as expected.

The $1/m$ HQET bare proper vertex has the form
\begin{eqnarray}
&&i\frac{C_k}{2m}gt^a\frac{1+\slash{v}}{2}\left[(1+\Lambda_k)(p+p')_\bot^\mu
+(\Lambda_{k0}+\Lambda_{k1}p_\bot^2+\Lambda'_{k1}p_\bot^{\prime2}
+\Lambda_{k2}q_\bot^2)v^\mu\right]
\nonumber\\
&&{}+i\frac{C_m}{4m}gt^a\frac{1+\slash{v}}{2}[\gamma^\mu,\slash{q}]
\frac{1+\slash{v}}{2}(1+\Lambda_m)\,,
\label{V1}
\end{eqnarray}
where all $\Lambda_i$ depend on $\omega$, $\Delta$;
$\Lambda'_{k1}(\omega,\Delta)=\Lambda_{k1}(\omega+\Delta,-\Delta)$;
$\Lambda_k(\omega,\Delta)=\Lambda_k(\omega+\Delta,-\Delta)$,
and similarly for $\Lambda_{k0}$, $\Lambda_{k2}$.
Similarly to the previous Section, we can see that
variation of the leading vertex function at $v\to v+\delta v$
coincides with that of the kinetic--energy vertex function at
$p_\bot\to p_\bot+\delta p_\bot$,
if $\delta v=\frac{C_k}{m}\delta p_\bot$.
This requires
\begin{equation}
\Lambda_k(\omega,\Delta)=\Lambda(\omega,\Delta)\,, \quad
\Lambda'_{k1}(\omega,\Delta)=
\frac{\partial\Lambda(\omega,\Delta)}{\partial\Delta}
\label{RIV}
\end{equation}
(and hence $\Lambda_{k1}(\omega,\Delta)=
\left(\frac{\partial}{\partial\omega}-\frac{\partial}{\partial\Delta}\right)
\Lambda(\omega,\Delta)$).
The Ward identities of Fig.~\ref{Fig:2} result in
\begin{equation}
\Lambda_{k0}(\omega,0)=-\frac{d\Sigma_{k0}(\omega)}{d\omega}\,, \quad
\Lambda_{k2}(\omega,0)=0
\label{WV}
\end{equation}
(in an abelian external field, $\Lambda_{k0}(\omega,\Delta)=
-\frac{\Sigma_{k0}(\omega+\Delta)-\Sigma_{k0}(\omega)}{\Delta}$,
$\Lambda_{k2}(\omega,\Delta)=0$).

Reparameterization invariance relates the spin--orbit vertex function
to the chromomagnetic one, but we shall not discuss details here.

The on--shell HQET vertex at the tree level is
\begin{equation}
\overline{u}_v(k') \left( v^\mu + C_k \frac{(k+k')^\mu}{2m}
+ C_m \frac{[\slash{q},\gamma^\mu]}{4m}
+ C_d \frac{q^2}{8m^2}v^\mu
+ C_s \frac{[\slash{k},\slash{q}]}{8m^2}v^\mu + \cdots \right) u_v(k)\,.
\label{HQETv}
\end{equation}
As we have demonstrated above,
there are no corrections to the first two terms.
Other terms have corrections starting from two loops,
if there is a finite--mass flavour (such as $c$ in $b$--quark HQET).
Expressing the on--shell QCD vertex via HQET spinors, we obtain
\begin{eqnarray}
&&\overline{u}_v(k') \Biggl[ \ep(q^2) \left( v^\mu + \frac{(k+k')^\mu}{2m}
- \frac{q^2+[\slash{k},\slash{q}]}{8m^2}v^\mu + \cdots \right)
\label{FF2}\\
&&\quad{}
+ \mu(q^2) \left( \frac{[\slash{q},\gamma^\mu]}{4m}
+ \frac{q^2+[\slash{k},\slash{q}]}{4m^2}v^\mu + \cdots \right) \Biggr] u_v(k)
\,.
\nonumber
\end{eqnarray}
Therefore, the coefficients in the HQET lagrangian are
\begin{equation}
C_k=1\,, \quad C_m=\mu\,, \quad C_d=8\ep'+2\mu-1\,, \quad C_s=2\mu-1\,.
\label{M2}
\end{equation}
The first one has no corrections~(\ref{Ck}).
The coefficients~(\ref{M2}) are not independent:
\begin{equation}
C_s=2C_m-1\,.
\label{RI2}
\end{equation}
Probably, reparameterization--invariance Ward identities yield
relations among corrections from finite--mass loops in HQET
which ensure the absence of corrections to~(\ref{RI2}).
However, we shall not trace details here.

Similarly, at the $1/m^3$ level, the coefficients in the HQET lagrangian are
\begin{equation}
C_{w1}=4\mu'+\tfrac{1}{2}\mu+\tfrac{1}{2}\,, \quad
C_{w2}=4\mu'+\tfrac{1}{2}\mu-\tfrac{1}{2}\,, \quad
C_{p'p}=\mu-1\,, \quad
C_M=-4\ep'-\tfrac{1}{2}\mu+\tfrac{1}{2}\,.
\label{M3}
\end{equation}
They are not independent:
\begin{equation}
C_{w2}=C_{w1}-1\,, \quad
C_{p'p}=C_m-1\,, \quad
C_M=\tfrac{1}{2}\left(C_m-C_d\right)\,.
\label{RI3}
\end{equation}
Calculation of $C_a$, $C_b$ requires matching amplitudes with two gluons.
Calculation of contact terms requires matching amplitudes with light quarks.

\section{Chromomagnetic interaction at two loops}

As we know, the kinetic coefficient $C_k(\mu)=1$,
and the only coefficient in the HQET lagrangian up to $1/m$ level
which is not known exactly is the chromomagnetic coefficient $V_m(\mu)$.
It is natural to find it from QCD/HQET matching at $\mu\sim m$
where no large logarithms appear.
Renormalization group can be used to obtain $C_m$ at $\mu\ll m$:
\begin{equation}
C_m(\mu) = C_m(m) \exp\left(-\int\limits_{\alpha_s(m)}^{\alpha_s(\mu)}
\frac{\gamma_m(\alpha)}{2\beta(\alpha)} \frac{d\alpha}{\alpha} \right)\,,
\label{RG}
\end{equation}
where $C_m(m)=1+C_1\frac{\alpha_s(m)}{4\pi}
+C_2\left(\frac{\alpha_s}{4\pi}\right)^2+\cdots$,
$\gamma_m=\frac{d\log Z_m}{d\log\mu}=\gamma_1\frac{\alpha_s}{4\pi}
+\gamma_2\left(\frac{\alpha_s}{4\pi}\right)^2+\cdots$
is the anomalous dimension of the chromomagnetic operator in HQET,
and the $\beta$--function is
$\beta=-\frac{1}{2}\frac{d\log\alpha_s}{d\log\mu}=\beta_1\frac{\alpha_s}{4\pi}
+\beta_2\left(\frac{\alpha_s}{4\pi}\right)^2+\cdots$
(where $\beta_1=\frac{11}{3}C_A-\frac{4}{3}T_F n_f$).
If $L=\log m/\mu$ is not very large,
it is better to retain all two--loop terms and neglect higher loops:
\begin{equation}
C_m(\mu) = 1 + \left(C_1 - \gamma_1 L \right) \frac{\alpha_s(m)}{4\pi}
+ \left[C_2 - \left(C_1\gamma_1+\gamma_2\right) L
+ \gamma_1\left(\gamma_1-\beta_1\right) L^2 \right]
\left(\frac{\alpha_s}{4\pi}\right)^2\,.
\label{RG1}
\end{equation}
This approximation holds up to relatively large $L$
because $C_2$ is numerically large.
If $L$ is parametrically large,
then it is better to sum leading and subleading logarithms:
\begin{equation}
C_m(\mu) =
\left(\frac{\alpha_s(\mu)}{\alpha_s(m)}\right)^{-\frac{\gamma_1}{2\beta_1}}
\left[ 1 + C_1 \frac{\alpha_s(m)}{4\pi}
 - \frac{\beta_1\gamma_2-\beta_2\gamma_1}{2\beta_1^2}
\frac{\alpha_s(\mu)-\alpha_s(m)}{4\pi} \right]\,.
\label{RG2}
\end{equation}
In this case, we cannot utilize $C_2$ without knowing $\gamma_3$.
In general, the solution of~(\ref{RG}) can be written as
\begin{equation}
C_m(\mu) = \hat{C}_m K(\mu)\,,\quad
\hat{C}_m = \alpha_s(m)^{\frac{\gamma_1}{2\beta_1}}(1+\delta c)\,,\quad
\delta c = c_1 \frac{\alpha_s(m)}{4\pi}
+ c_2 \left(\frac{\alpha_s(m)}{4\pi}\right)^2+\cdots
\label{RG3}
\end{equation}
where $\hat{C}_m$ is scale-- and scheme--independent.

As a simple application,
we consider $B$--$B^*$ mass splitting~\cite{Mannel,BSUV}%
\footnote{in~\cite{Mannel}, $\rho_{mm}^3$ is missing;
in~\cite{BSUV}, the leading logarithmic running of $C_m(\mu)$
has a wrong sign.}
\begin{equation}
m_{B^*}-m_B = \frac{2C_m(\mu)}{3m}\mu_m^2(\mu) + \frac{1}{3m^2}
\left[ C_m(\mu) \rho_{km}^3(\mu) + C_m^2(\mu) \rho_{mm}^3(\mu)
- C_s(\mu) \rho_s^3(\mu) \right]\,,
\label{spl}
\end{equation}
where $\mu_m^2(\mu)$ and $\rho_s^3(\mu)$ are local matrix elements
of chromomagnetic interaction and spin--orbit one,
while $\rho_{km}^3(\mu)$ and $\rho_{mm}^3(\mu)$ are kinetic--chromomagnetic
and chromomagnetic--chromomagnetic bilocal matrix elements
(in the later case, there are two $\gamma$--matrix structures,
1 and $\sigma_{\mu\nu}$; the coefficient of the second one is implied here).
Introducing renormalization group invariants
\begin{eqnarray}
&&\hat{\mu}_m^2 = K(\mu) \mu_m^2(\mu)\,,\quad
\hat{\rho}_{km}^3 =
K(\mu) \rho_{km}^3(\mu) + \left[1-K(\mu)\right] \rho_s^3(\mu) \,,\quad
\nonumber\\
&&\hat{\rho}_{mm}^3 = K^2(\mu) \rho_{mm}^3 \,,\quad
\hat{\rho}_s^3 = \rho_s^3(\mu)\,,
\label{spl2}
\end{eqnarray}
we can rewrite it as
\begin{equation}
m_{B^*}-m_B = \frac{2\hat{C}_m}{3m} \hat{\mu}_m^2
+ \frac{1}{3m^2} \left[
\hat{C}_m \left(\hat{\rho}_{km}^3-2\hat{\rho}_s^3\right)
+ \hat{C}_m^2 \hat{\rho}_{mm}^3 + \hat{\rho}_s^3 \right]\,.
\label{spl3}
\end{equation}

\begin{figure}[p]
\includegraphics[width=0.975\linewidth]{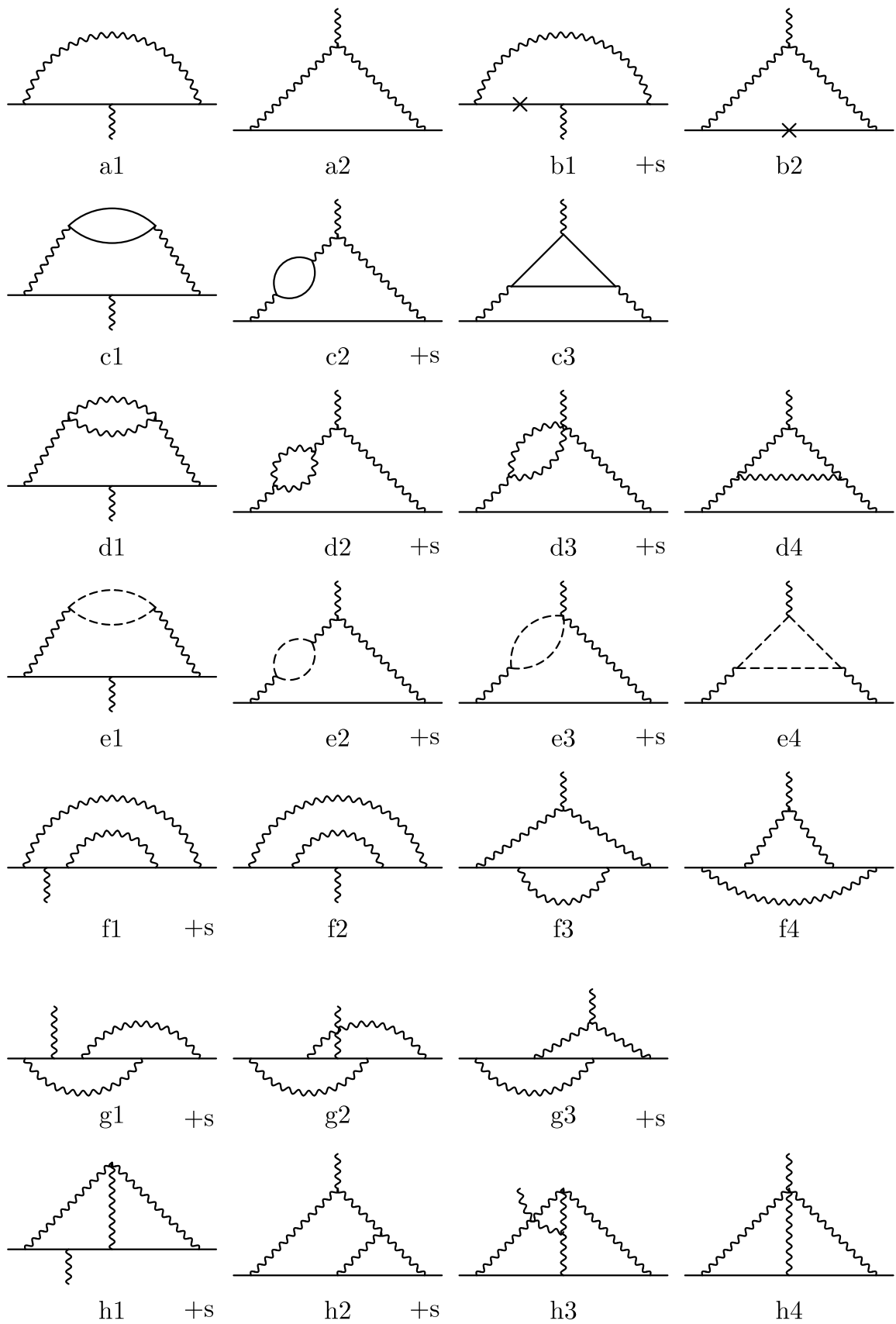}
\caption{Diagrams for the QCD proper vertex}
\label{Fig:3}
\end{figure}

In order to obtain $C_m$, we should calculate the heavy--quark
chromomagnetic moment $\mu$ (Fig.~\ref{Fig:3}).
All on--shell massive integrals can be reduced to 3 basis ones
\begin{equation}
I_0^2 = \raisebox{-1cm}{\includegraphics{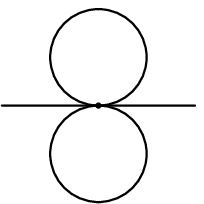}} ,\quad
I_1 = \raisebox{-0.6cm}{\includegraphics{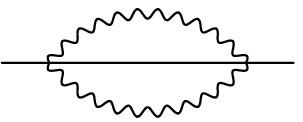}} ,\quad
I_2 = \raisebox{-0.6cm}{\includegraphics{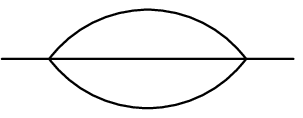}}
\label{I012}
\end{equation}
using integration by parts~\cite{GBGS}--\cite{Broadhurst}.
$I_0^2$ and $I_1$ are expressed via $\Gamma$--functions of $d$;
$I_2$ is expressed via $I_0^2$, $I_1$,
and one difficult convergent integral~\cite{Broadhurst}
\begin{equation}
I=\pi^2\log 2-\frac{3}{2}\zeta(3)+O(\ep)\,.
\label{I2}
\end{equation}
The result has the structure
\begin{eqnarray}
&&\mu = 1 + \frac{g_0^2 m^{-2\ep}}{(4\pi)^{d/2}} (C_F,C_A) \times I_0
\label{mu}\\
&&\quad{} + \frac{g_0^4 m^{-4\ep}}{(4\pi)^d}
(C_F^2,C_F C_A,C_A^2,C_F T_F n_l,C_A T_F n_l,C_F T_F,C_A T_F)
\times (I_0^2,I_1,I_2)\,.
\nonumber
\end{eqnarray}
Now we express it via $\alpha_s(\mu)$ and expand in $\ep$.
The coefficient of $1/\ep$ gives the anomalous dimension
\begin{equation}
\gamma_m = 2 C_A \frac{\alpha_s}{4\pi}
+ \frac{4}{9} C_A \left(17C_A-13T_F n_f\right)
\left(\frac{\alpha_s}{4\pi}\right)^2
+ \cdots
\label{gam}
\end{equation}
The chromomagnetic interaction coefficient at $\mu=m$ is
\begin{eqnarray}
&&\hspace{-6mm}
C_m(m) = 1 + 2(C_F+C_A) \frac{\alpha_s(m)}{4\pi}
\nonumber\\&&\hspace{-6mm}
+ \Biggl[ C_F^2 \left(-8I+\frac{20}{3}\pi^2-31\right)
+ C_F C_A \left(\frac{4}{3}I+\frac{4}{3}\pi^2+\frac{269}{9}\right)
+ C_A^2 \left(\frac{4}{3}I-\frac{17}{9}\pi^2+\frac{805}{27}\right)
\nonumber\\&&\hspace{-1mm}
+ C_F T_F n_l \left(-\frac{100}{9}\right)
+ C_A T_F n_l \left(-\frac{4}{9}\pi^2-\frac{299}{27}\right)
\label{Cm}\\&&\hspace{-1mm}
+ C_F T_F \left(-\frac{16}{3}\pi^2+\frac{476}{9}\right)
+ C_A T_F \left(\pi^2-\frac{298}{27}\right)
\Biggr] \left(\frac{\alpha_s}{4\pi}\right)^2
\nonumber\\
&&\hspace{-6mm} = 1 + \frac{13}{6} \frac{\alpha_s(m)}{\pi} +
\left( 21.79 - 1.91 n_l \right) \left(\frac{\alpha_s}{\pi}\right)^2\,.
\nonumber
\end{eqnarray}
The coefficient of $(\alpha_s/\pi)^2$ is about 11 for $n_l=4$ light flavours.
It is 40\% less than the expectation
based on naive nonabelianization~\cite{BG}.
The contribution of the heavy quark loop to this coefficient is merely $-0.1$.

\section{Chromomagnetic interaction at higher loops}

Perturbation series for $C_m$ can be rewritten via $\beta_1$
instead of $n_f$:
\begin{equation}
C_m(\mu) = 1 + \sum_{L=1}^{\infty} \sum_{n=0}^{L-1} a_{Ln} \beta_1^n \alpha_s^L
= 1 + \frac{1}{\beta_1} f(\beta_1 \alpha_s) + O\left(\frac{1}{\beta_1^2}\right)
\,.
\label{pert}
\end{equation}
There is no sensible limit of QCD in which $\beta_1$ may be considered
a large parameter (except, may be, $n_f\to-\infty$).
However, retaining only the leading $\beta_1$ terms often gives
a good approximation to exact multi--loop results~\cite{BG}.
This limit is believed to provide information about summability
of perturbation series~\cite{Mueller}.
At the first order in $1/\beta_1$, multiplicative renormalization
amounts to subtraction of $1/\ep^n$ terms;
\begin{equation}
\frac{\beta_1 g_0^2}{(4\pi)^2} = \bar{\mu}^{2\ep} \frac{\beta}{1+\beta/\ep}\,,
\quad \beta=\frac{\beta_1 \alpha_s}{4\pi}=\frac{1}{2\log\mu/\LMS}\,.
\label{beta}
\end{equation}
The perturbation series~(\ref{pert}) can be rewritten as
\begin{equation}
C_m(\mu) = 1 + \frac{1}{\beta_1} \sum_{L=1}^{\infty}
\frac{F(\ep,L\ep)}{L} \left(\frac{\beta}{\ep+\beta}\right)^L
- \mathrm{(subtractions)} + O\left(\frac{1}{\beta_1^2}\right)\,.
\label{pert2}
\end{equation}

Knowledge of the function $F(\ep,u)$ allows one to obtain
the anomalous dimension
\begin{equation}
\gamma_m = \frac{2\beta}{\beta_1} F(-\beta,0)
+ O\left(\frac{1}{\beta_1^2}\right)
\label{rgam}
\end{equation}
and the finite term
\begin{equation}
C_m(\mu) = 1 + \frac{1}{\beta_1} \int\limits_{-\beta}^{0} d\ep
\frac{F(\ep,0)-F(0,0)}{\ep}
+ \frac{1}{\beta_1} \int\limits_{0}^{\infty} du\, e^{-u/\beta}
\frac{F(0,u)-F(0,0)}{u} + O\left(\frac{1}{\beta_1^2}\right)
\label{rCm}
\end{equation}
(this method was used in~\cite{BG}; see references in this paper).
Renormalization group invariant~(\ref{RG3}) is
\begin{equation}
\delta c = \frac{1}{\beta_1} \int_0^\infty du\,
e^{-\frac{4\pi}{\beta_1\alpha_s}u}S(u)
+ O\left(\frac{1}{\beta_1^2}\right)\,,\quad
S(u) = e^{-\frac{5}{3}u} \left. \frac{F(0,u)-F(0,0)}{u} \right|_{\mu=m}
\label{rCm2}
\end{equation}
(here $\alpha_s$ is taken at $\mu=m$ in the $V$--scheme,
$\exp\bigl(-\frac{4\pi}{\beta_1\alpha_s}u\bigr)
=\bigl(\frac{\Lambda_V}{m}\bigr)^{-2u}$).

\begin{figure}[ht]
\includegraphics[width=\linewidth]{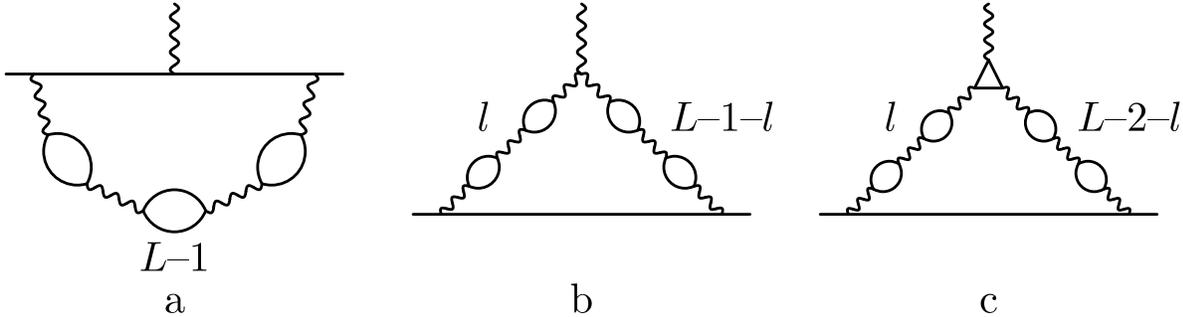}
\caption{$L$--loop diagrams with the maximum number of quark loops.}
\label{Fig:4}
\end{figure}

The function $F(\ep,u)$ is determined by the coefficient of the highest
degree of $n_f$ in the $L$--loop term, which is given by the diagrams
in Fig.~\ref{Fig:4}.
Calculating them, we obtain
\begin{eqnarray}
&&\hspace{-6mm}
F(\ep,u) = \left(\frac{\mu}{m}\right)^{2u}
e^{\gamma\ep} \frac{\Gamma(1+u)\Gamma(1-2u)}{\Gamma(3-u-\ep)}
D(\ep)^{u/\ep-1} N(\ep,u)
\nonumber\\
&&\hspace{-6mm}
D(\ep) = 6 e^{\gamma\ep} \Gamma(1+\ep) B(2-\ep,2-\ep) =
1 + {\textstyle\frac{5}{3}} \ep + \cdots
\label{Feu}\\
&&\hspace{-6mm}
N(\ep,u) = C_F 4u(1+u-2\ep u)
+ C_A \frac{2-u-\ep}{2(1-\ep)} (2+3u-5\ep-6\ep u+2\ep^2+4\ep^2 u)\,.
\nonumber
\end{eqnarray}
This gives the anomalous dimension
\begin{eqnarray}
&&\hspace{-6mm}
\gamma_m = C_A \frac{\alpha_s}{2\pi}
\frac{\beta(1+2\beta)\Gamma(5+2\beta)}
{24(1+\beta)\Gamma^3(2+\beta)\Gamma(1-\beta)}
\label{rgam2}\\
&&\hspace{-4mm}\quad{}
= C_A \frac{\alpha_s}{2\pi} \left[1
+ \frac{13}{6} \frac{\beta_1 \alpha_s}{4\pi}
- \frac{1}{2} \left(\frac{\beta_1 \alpha_s}{4\pi}\right)^2
+ \cdots \right]\,.
\nonumber
\end{eqnarray}
This perturbation series is convergent with the radius
$\beta_1|\alpha_s|<4\pi$.
The Borel image of $\delta c$
\begin{equation}
S(u) = \frac{\Gamma(u)\Gamma(1-2u)}{\Gamma(3-u)} \left[ 4u(1+u)C_F
+ \tfrac{1}{2}(2-u)(2+3u)C_A \right] - e^{-\frac{5}{3}u}\frac{C_A}{u}
\label{Su}
\end{equation}
has infrared renormalon poles at $u=\frac{n}{2}$.
They produce ambiguities in the sum of the perturbation series for $\delta c$,
which are of order of the residues ${}\sim(\Lambda_V/m)^n$.
The leading ambiguity ($u=\frac{1}{2}$) is
\begin{equation}
\Delta \hat{C}_m =
\left(1+\frac{7}{8}\frac{C_A}{C_F}\right)\frac{\Delta m}{m}\,,
\label{dCm}
\end{equation}
where $\Delta m$ is the ambiguity of the heavy--quark
pole mass~\cite{BB,BSUV2}.

Physical quantities, such as the mass splitting~(\ref{spl}),
are factorized into short--distance coefficients
and long--distance hadronic matrix elements.
In regularization schemes without a hard momentum cut--off,
such as $\overline{\mathrm{MS}}$, Wilson coefficients
also contain large--distance contributions which produce
infrared renormalon ambiguities.
Likewise, hadronic matrix elements contain small--distance
contributions which produce ultraviolet renormalon ambiguities.
In other words, the separation into short-- and long--distance
contributions is ambiguous;
only when they are combined to form a physical quantity,
an unambiguous result is obtained.
Cancellations between infrared and ultraviolet renormalon
ambiguities in HQET were traced in~\cite{NS}.

\begin{figure}[ht]
\includegraphics[width=\linewidth]{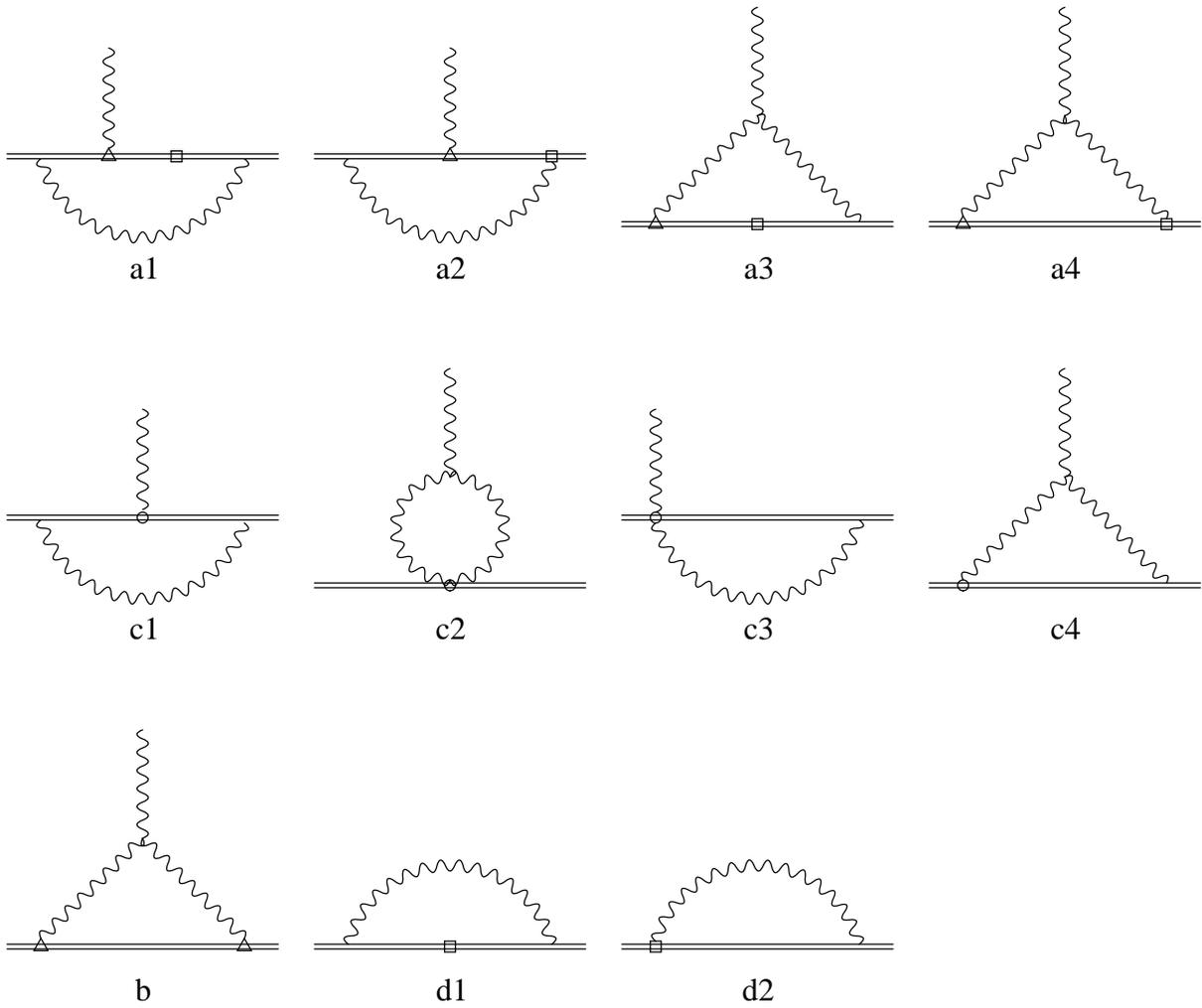}
\caption{Diagrams for $\rho_i^3$; quark loops are inserted in all
possible ways.}
\label{Fig:5}
\end{figure}

Ultraviolet renormalon ambiguities in matrix elements $\rho_i^3$
don't depend on external states, and may be calculated
at the level of quarks and gluons (Fig.~\ref{Fig:5}).
Note that there is an ultraviolet renormalon ambiguity
in the wave function renormalization
$\Delta Z_Q=\frac{3}{2}\frac{\Delta m}{m}$ (Fig.~\ref{Fig:5}d).
The result is
\begin{equation}
\Delta\rho_{km}^3=-\frac{2}{3}\frac{C_A}{C_F}\mu_m^2\Delta m\,,\quad
\Delta\rho_{mm}^3=-\frac{19}{12}\frac{C_A}{C_F}\mu_m^2\Delta m\,,\quad
\Delta\rho_s^3=-\frac{1}{2}\frac{C_A}{C_F}\mu_m^2\Delta m\,.
\end{equation}
The sum of ultraviolet ambiguities of the $1/m^2$
contributions to~(\ref{spl}) cancels the infrared ambiguity
of the leading term.

The requirement of cancellation of renormalon ambiguities
in the mass splitting~(\ref{spl2}) for all $m$ allows us
to establish the structure of the leading infrared renormalon
singularity in $S(u)$ at $u=\frac{1}{2}$ beyond the large $\beta_1$ limit.
The ultraviolet ambiguity of the square bracket in~(\ref{spl2})
should be equal to $\hat{\mu}_m^2$ times
\begin{equation}
\Lambda_V=m\,e^{-\frac{2\pi}{\beta_1\alpha_s}}
\alpha_s^{-\frac{\beta_2}{2\beta_1^2}}[1+O(\alpha_s)]\,.
\label{Lam}
\end{equation}
In order to reproduce the correct fractional powers of $\alpha_s$,
$S(u)$ in~(\ref{rCm2}) should have the branch point at $u=\frac{1}{2}$
instead of a pole:
\begin{equation}
S(u)=\frac{1}{\left(\frac{1}{2}-u\right)^{1+\beta_2/2\beta_1^2}}
\left[ 2 C_F K_1 - \frac{1}{3} C_A K_2
+ \frac{19}{12} \frac{C_A K_3}{\left(\frac{1}{2}-u\right)^{-\gamma_1/2\beta_1}}
+ \frac{1}{2} \frac{C_A K_4}{\left(\frac{1}{2}-u\right)^{\gamma_1/2\beta_1}}
\right]\,,
\end{equation}
where omitted terms are suppressed as $\frac{1}{2}-u$ compared to
the displayed ones.
Normalization constants are known in the large $\beta_1$ limit only:
$K_i=1+O(1/\beta_1)$.
The large--order behaviour of the perturbation series
for $\delta c$ is
\begin{equation}
c_{n+1} = n!\,(2\beta_1)^n\,n^{\beta_2/2\beta_1^2}\,
\left[ 4 C_F K_1 - \tfrac{2}{3} C_A K_2 
+ \tfrac{19}{6} C_A K_3 n^{-\gamma_1/2\beta_1}
+ C_A K_4 n^{\gamma_1/2\beta_1} \right]\,,
\end{equation}
where omitted terms are suppressed as $1/n$ compared to the displayed ones.

\textbf{Acknowledgements}. I am grateful to A.~Czarnecki and M.~Neubert
for collaboration in writing~\cite{CG,GN};
to S.~Groote for ongoing collaboration;
to C.~Balzereit for discussing~\cite{Balzereit,Balzereit2};
to T.~Mannel for useful discussions;
to J.~G.~K\"orner for hospitality at Mainz during preparation of this talk;
and to M.~Beyer for organization of the workshop.

\end{document}